\documentclass{iau} 
\usepackage{graphicx}

\title[Magnetic white dwarfs] 
{Clues to the origin and properties of magnetic white dwarfs}

\author[A. Kawka]   
{Adela Kawka$^1$}

\affiliation{$^1$International Centre for Radio Astronomy Research - Curtin University\\
GPO Box U1987, Perth, WA 6845, Australia\\ email: {\tt adela.kawka@curtin.edu.au}}

\pubyear{2019}
\volume{357}  
\jname{White dwarfs as probes of fundamental physics and tracers of planetary, stellar \& galactic evolution}
\editors{A.C. Editor, B.D. Editor \& C.E. Editor, eds.}
\begin{document}

\maketitle

\begin{abstract}
A significant fraction of white dwarfs possess a magnetic field with strengths
ranging from a few kG up to about 1000 MG. However, the incidence of magnetism
varies when the white dwarf population is broken down into different spectral
types providing clues on the formation of magnetic fields in white dwarfs.
Several scenarios for the origin of magnetic fields
have been proposed from a fossil field origin to dynamo generation at various stages of evolution. Offset dipoles are often assumed sufficient to model
the field structure, however time-resolved spectropolarimetric observations have revealed more complex structures such as magnetic spots or multipoles.
Surface mapping of these field structures combined with measured rotation rates help distinguish scenarios involving single star evolution from other scenarios involving binary interactions.
I describe key observational properties of magnetic
white dwarfs such as age, mass, and field strength, and confront proposed formation scenarios with these properties.

\end{abstract}

\firstsection 
\section{Introduction}

Following the discovery of a magnetic field in 78 Virginis 
(\cite[Babcock 1947]{bab1947}), \cite[Blackett (1947)]{bla1947} contemplated 
the presence of stronger magnetic fields in white dwarfs assuming magnetic 
flux conservation throughout evolution. The strong circular 
polarization spectrum of the first known magnetic white dwarf, 
Grw+70$^\circ$\,8247 (\cite[Kemp et al. 1970]{kem1970}), implied a
longitudinal field measurement of $\approx10^7$ G, which was nearly a thousand 
times the field strength observed in Ap stars at the time. Following the 
original discovery, many more magnetic white dwarfs were found with strong 
magnetic fields ($B \gtrsim 10^7$ G) including G195-19 as the prototype for 
the cool, continuum-like, helium-rich magnetic white dwarfs 
(DCP, \cite[Angel \& Landstreet 1971]{ang1971}) and G99-37 as the prototype 
for carbon-enriched magnetic white dwarfs (DQP but originally classified as DGp,
\cite[Greenstein 1969]{gre1969}, \cite[Landstreet \& Angel 1971]{lan1971}). 
Early observations already demonstrated that magnetic fields are present 
across all spectral types. Although a search for lower strength magnetic fields 
($\lesssim 5$~MG) carried out by \cite[Angel, Borra \& Landstreet (1981)]{ang1981} proved negative, 
the number of white dwarfs with magnetic fields greater than 1~MG steadily 
grew by 1995 to 40 objects (see \cite[Schmidt \& Smith 1995]{sch1995}). 
Spectropolarimetric measurements proved capable of reaching below 1~MG field 
strength with the discovery of two low-field white dwarfs (0.1 MG) in a large 
survey of 170 objects (\cite[Schmidt \& Smith 1994,1995]{sch1994,sch1995}). 
With even larger surveys, such as the Sloan Digital Sky Survey (SDSS) and with 
larger telescopes and more sensitive instruments, the number of
magnetic white dwarfs has grown to over 600 
(see, e.g., \cite[Ferrario, de Martino \& G\"{a}nsicke 2015]{fer2015}) but 
still including only 18 objects with fields below 1~MG (\cite[Kawka \& Vennes 2012]{kaw2012}, \cite[Landstreet, et al. 2017]{lan2017}).

This paper aims to summarise our current understanding of the magnetic white dwarf population and the
methods used to detect and determine magnetic fields in white dwarfs (Section 2). The properties of the magnetic white dwarf 
population are presented in Section 3 and the various models for the origin of magnetic fields in white dwarfs are presented
in Section 4. Finally, Section 5 exposes current challenges in modelling magnetic atmospheres 
in particular the effect of magnetic fields on convective motion and the atmospheric structure.

\section{Measuring magnetic field strengths}

\begin{figure}
    \centering
    \includegraphics[viewport=1 1 575 575,clip, width=0.8\textwidth]{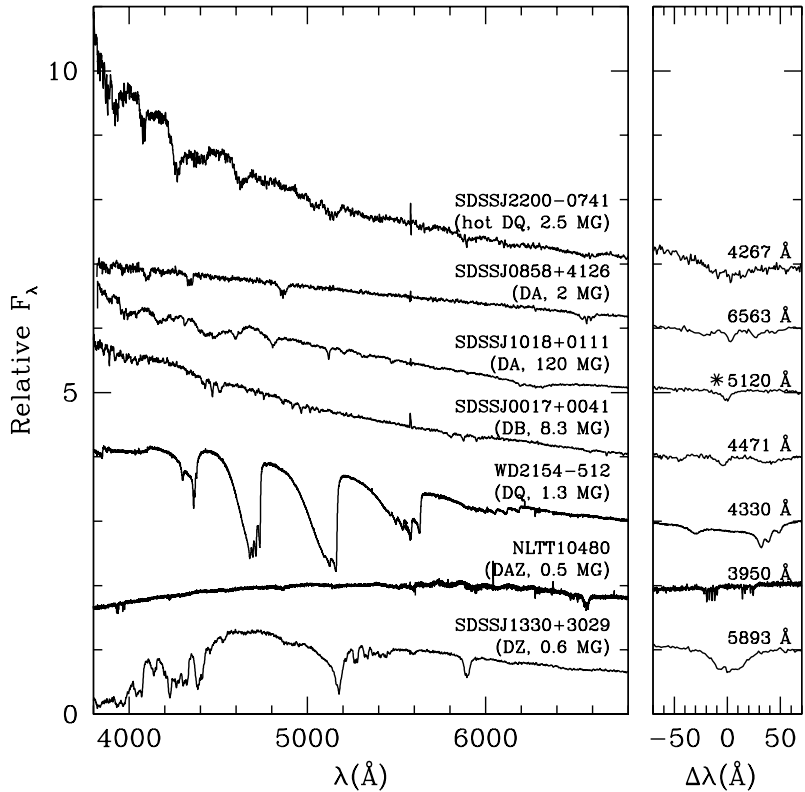}
    \caption{Example spectra showing the different spectral classes for which 
magnetism has been detected. The left panel shows the optical spectra. The left 
panel highlights a particular feature that is representative of the spectral 
type which is centred on the given wavelength. The absorption line at 5120 \AA,
marked with a star, in SDSSJ1018$+$0111 is an example of a stationary feature.
In this case it is a Zeeman component of H$\beta$.}
    \label{fig_spectra}
\end{figure}

There are two complementary ways to detect and measure magnetic fields. The most straight forward 
follows the detection of Zeeman splitted lines. Depending on the atomic structure and magnetic 
field strength, the Zeeman effect can be calculated using distinct methods. Fig.~\ref{fig_spectra}
shows examples of the different classes of magnetic white dwarfs. The magnetic DC white dwarfs, which have featureless 
spectra, are not included in the plot. The magnetic DQ shown is WD~2154-512 which exhibits CH molecular bands in its spectrum (shown in the left 
panel of Fig.~\ref{fig_spectra} and centred at 4330 \AA) and which were used in measuring its
magnetic field strength from spectropolarimetry (\cite[Berdyugina et al. 2007]{ber2007}). The figure also shows the different regimes of
the Zeeman effect on various elements.

The normal Zeeman effect occurs when a spectral line of an atom with zero spin ($S=0$) splits into three components when immersed in a magnetic field. In this case, we can calculate the magnetic field strength using:

\begin{equation}
\label{eqn_norm_zeeman}
\Delta \lambda = \frac{eB\lambda^2}{4\pi m_e c^2} \approx 4.67\times10^{-7} \lambda^2 B
\end{equation}

where $\lambda$ is the wavelength in \AA, $B$ is the magnetic field in MG, $e$ is the electron charge (e.s.u.), $m_e$ is the electron
rest mass, and $c$ is the speed of light.

The anomalous Zeeman effect depends on the electron spin and occurs in atoms with non-zero spin. The splitting in this regime can
be complex and is observed at lower field strength where the spin and orbital angular momenta remain coupled. In general, the energy
level of an electron can be described by three quantum numbers, $J$ which is the total angular momentum, $L$ which is the
orbital angular momentum, and $S$ which is the spin orbital momentum. When an atom is placed in a magnetic field its levels
split into $2J+1$ components. These are defined by the magnetic quantum number $m= -J, -J+1, ..., J-1, J$ and the Zeeman shifted
lines can be calculated using:

\begin{equation}
\Delta \lambda = \frac{eB\lambda^2}{4\pi m_e c^2}(g_l m_l - g_u m_u) \approx 4.67\times10^{-7} \lambda^2 B (g_l m_l - g_u m_u)
\end{equation}

where $m_u$ and $m_l$ are the magnetic quantum numbers of the upper and lower levels, respectively.
The permitted transitions are defined by $\Delta m = 0,\pm1$. Here, $\Delta m = 0$ defines the $\pi$ components
and $\Delta m = \pm1$ the $\sigma$ components. The Land\'e factors for the upper and lower levels are $g_u$ and $g_l$, respectively 
and they can be calculated assuming LS coupling:

\begin{equation}
g = 1+\frac{J(J+1)-L(L-1)+S(S+1)}{2J(J+1)},
\end{equation}

The LS coupling is generally valid for lighter atoms such as sodium, magnesium, aluminium and calcium. For heavier atoms
like iron LS coupling is no longer valid and other coupling needs to be applied. The Vienna Atomic Line Database\footnote{http://vald.astro.uu.se/ \cite[(Ryabchikova et al. 2015)]{rya2015}.} provides atomic data including Land\'e factors for light and heavy elements.

The Paschen-Back effect occurs when the $m_l$ degeneracy is removed. In this 
regime the spin and orbital angular momenta decouple and the 
anomalous Zeeman effect reverts to the normal or linear Zeeman effect and 
Eqn.~\ref{eqn_norm_zeeman} can be used. The linear
Zeeman splitting of hydrogen atoms falls into this regime.

As the magnetic field strength increases, the quadratic effect begins to take 
over. In this regime, the $l$ degeneracy is also removed and shifts in the 
$\pi$ components and asymmetry in the line profiles are observed. Increasing 
the field strength will cause even more complex Zeeman patterns and for more 
detail on these higher field regimes see \cite[Ferrario et al. (2019)]{fer2019}.
\cite[Landi Degl'Innocenti \& Landolfi (2004)]{lan2004} present conveniently 
written Zeeman splitting formulae for the low field regime and quadratic 
effect. For further reading consult \cite[Herzberg (1945)]{her1945}. 
Table~\ref{tbl_zeeman} lists references to available calculations of line 
shifts and transition probabilities for hydrogen and helium at stronger 
magnetic fields, as well as some heavier elements such as sodium, calcium and 
carbon. For helium, the calculations of \cite[Becken \& Schmelcher (2001)]{bec2001} and \cite[Al-Hujaj \& Schmelcher (2003)]{alh2003}
supersede the calculations of \cite[Kemic (1974)]{kem1974} and allowed the 
measurement of magnetic field strengths of strongly magnetic
DB white dwarfs (\cite[Jordan et al. 1998, Wickramasinghe et al. 2002]{jor1998,wic2002}). 
Measuring magnetic field strengths at high fields ($\gtrsim 100$~MG) can be 
difficult due to the complexity of the Zeeman pattern. Also, the magnetic 
field on a white dwarf surface is not uniform and can vary significantly, up 
to a factor of two even if a simple dipole is assumed. This can smear out most 
of the absorption features, however some line components become stationary 
where the wavelength goes through a minimum or maximum as a function of the
field strength. These stationary components can be observed in spectra even 
when the field varies considerably (see Fig~\ref{fig_spectra}).

\begin{table}
  \begin{center}
  \caption{Calculations of Zeeman splitted lines.}
  \label{tbl_zeeman}
 {\scriptsize
  \begin{tabular}{lccl}\hline 
{\bf Element} & {\bf $B$ Range (G)} & {\bf Spectral range} & {\bf Reference} \\ 
\hline
Hydrogen   & $0 - 5\times10^{12}$    & Lyman, Balmer, Paschen, Brackett & \cite[Schimeczek \& Wunner (2014)]{sch2014} \\ 
Helium     & $0 - 2.35\times10^{11}$ &      HeI                         & \cite[Becken \& Schmelcher (2001)]{bec2001} \\
Helium     & $2.35\times10^{11} - 2.35\times10^{13}$ & HeI              & \cite[Al-Hujaj \& Schmelcher (2003)]{alh2003} \\
Sodium     & $0 - 4.7\times10^7$     & $n=3-7$                          & \cite[Gonz{\'a}lez-F{\'e}rez \& Schmelcher (2003)]{gon2003} \\
Calcium    & $10^6 - 10^8$           & CaII H\&K                        & \cite[Kemic (1975)]{kem1975} \\
\hline 
  \end{tabular}
  }
 \end{center}
\end{table}

Spectropolarimetric measurements add additional geometric information following a magnetic field detection. 
To date, most spectropolarimetric observations have measured
the circular polarization, the Stokes parameter $V$. The longitudinal field strength ($B_l$ in G) can be determined using:
\begin{equation}
    v = \frac{V}{I} = \frac{dI}{d\lambda}\frac{4.67 \times 10^{-13} \lambda^2 B_l}{I}
\end{equation}

\begin{figure}
    \centering
    \includegraphics[viewport=1 302 570 555,clip, width=0.53\textwidth]{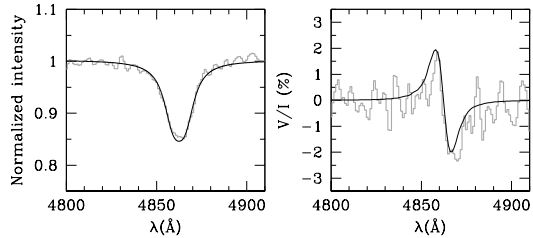}
    \caption{Flux (left) and circular polarization (right) spectra of NLTT~2219. The flux
    spectrum (grey) is compared to the best-fitting H$\beta$ line profile (black). The circular
    polarization spectrum is compared to a model polarization spectrum at $B_l = -97$~kG (\cite[Kawka \& Vennes 2012]{kaw2012}).}
    \label{fig_spol}
\end{figure}

This equation provides for a determination of the longitudinal field strength from the measured polarization $V$ and the intensity
spectrum $I$ for stars where the Zeeman splitting is smaller than the width of the line profile. Several low field
DA white dwarfs have been recently detected using this technique (\cite[e.g., Kawka \& Vennes 2012, Landstreet \& Bagnulo 2019]{kaw2012,lan2019a,lan2019b}). Fig.~\ref{fig_spol} shows
the H$\beta$ line profile of the weakly magnetic NLTT~2219 as well as the circular polarization spectrum
showing the polarized $\sigma$ components.

Spectropolarimetry has also been used to detect and measure magnetic fields in 
the carbon polluted DQ white dwarfs. The first DQ white dwarf
that showed circular polarization was G99-37 
(\cite[Angel \& Landstreet 1974]{ang1974}) followed by LP~790-29 
(\cite[Liebert et al. 1978]{lie1978}). Estimating a magnetic field strength is 
still difficult for DQ white dwarfs because of 
their relatively broad molecular carbon bands. However, \cite[Angel \& Landstreet (1974)]{ang1974} were able to obtain an estimate
of the magnetic field of G99-37 using CH molecular bands, a rare feature among 
DQ white dwarfs. \cite[Berdyugina et al. (2007)]{ber2007} 
and \cite[Vornanen et al. (2010)]{vor2010} further developed the method 
involving CH molecular bands to measure magnetic 
fields and determined a longitudinal field strength of $7.3\pm0.3$~MG for 
G99-37. The same method enabled them to 
identify the second magnetic DQ with CH bands (GJ~841B) with a field strength 
$B_l = 1.3\pm0.5$~MG.

Currently, there are several instruments capable of spectropolarimetry. These include the FOcal Reducer and low dispersion Spectrograph
(FORS2) at the European Southern Observatory (\cite[Appenzeller, et al. 1998]{app1998}) located at the Paranal Observatory in Chile, 
the Echelle SpectroPolarimetric Device for the Observation of Stars (\cite[ESPaDOnS: Donati et al. 2006)]{don2006}) 
on the Canada-France-Hawaii Telescope (CFHT) located on
Mauna Kea in Hawaii, the Intermediate-dispersion Spectrograph and Imaging System (ISIS) on the William Herschel Telescope (WHT) located
on La Palma in the Canary islands, Spain and the Robert Stobie Prime Focus Imaging Spectrograph (RSS) on the South African Large
Telescope near Sutherland in South Africa.

\section{Population properties}

\begin{figure}
    \centering
    \includegraphics[viewport=1 25 565 567,clip, width=0.5\textwidth]{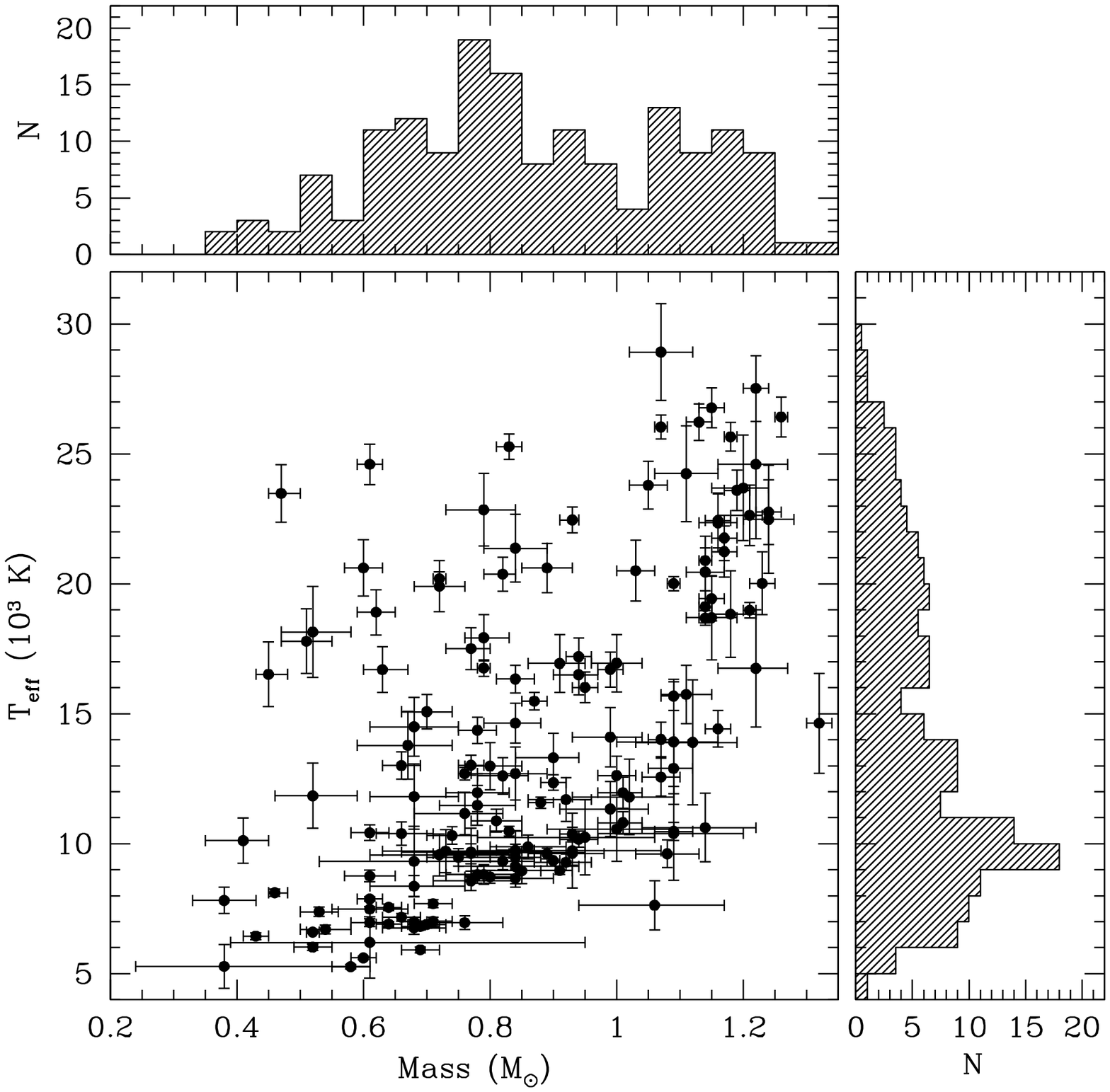}%
    \includegraphics[viewport=1 25 565 567,clip, width=0.5\textwidth]{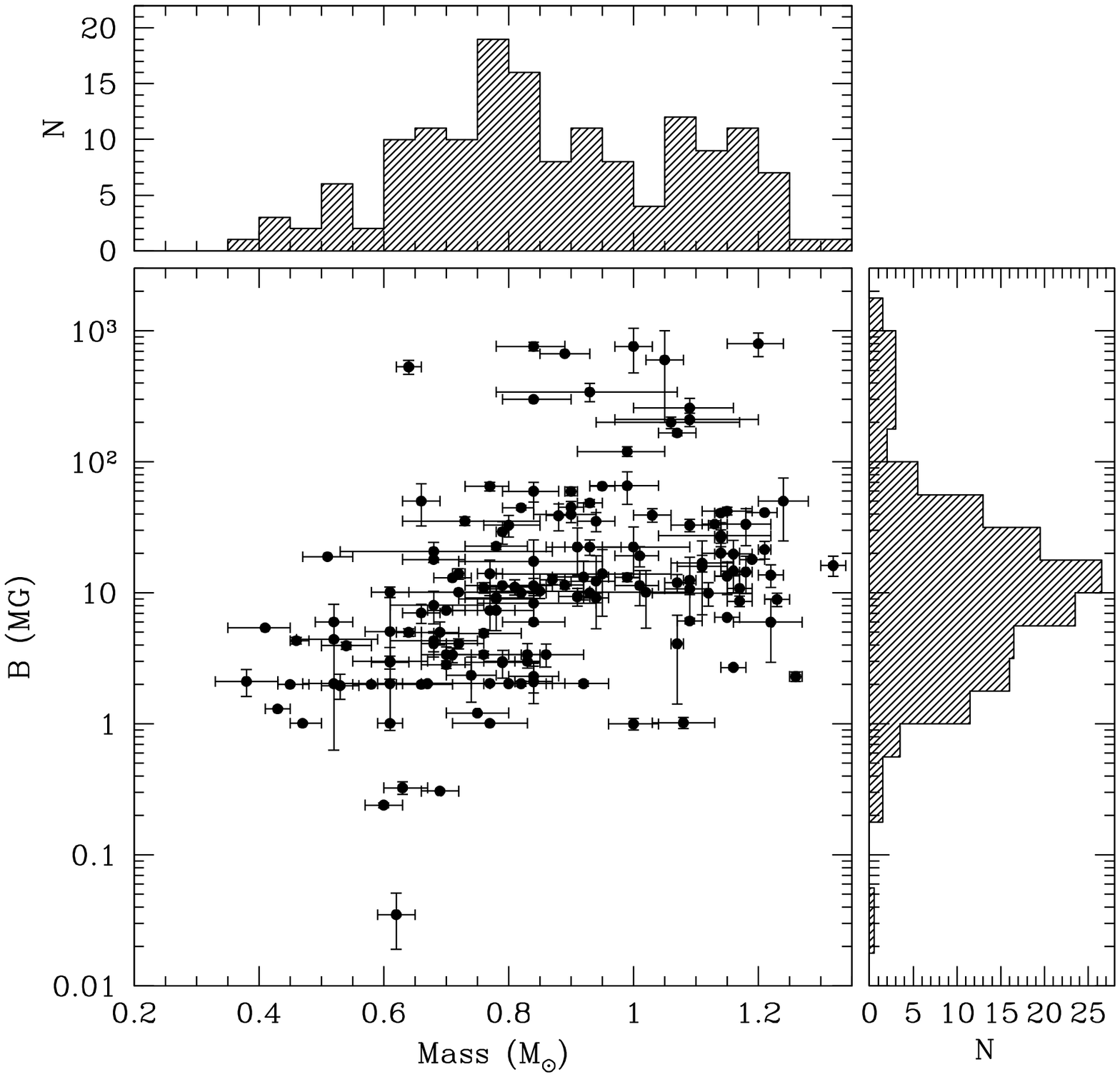}
    \caption{Mass versus temperature distribution (left) and mass versus magnetic field strength distribution (right) of magnetic DA white dwarfs from SDSS DR12.}
    \label{fig_mass_teff_dist}
\end{figure}

To investigate the properties of magnetic white dwarfs as a population, the 
sample of known magnetic DA white dwarfs was cross-correlated with SDSS DR12 
photometric measurements (\cite[Alam et al. 2015]{ala2015}) and Gaia DR2
parallax measurements (\cite[Lindegren et al. 2018]{lin2018}). The effective 
temperature and mass were determined by fitting the SDSS
photometric measurements with synthetic magnitudes that were set to the 
stellar distance using Gaia parallaxes and the stellar radius $R$ by applying 
mass-radius relations from \cite[Benvenuto \& Althaus (1999)]{ben1999}.
Fig.~\ref{fig_mass_teff_dist} shows the distribution of temperature and 
magnetic field strength versus mass for a sample of known magnetic white 
dwarfs with available SDSS photometric measurements and Gaia parallaxes.
The mass distribution shows that there are two peaks with the main one at 
$\sim 0.8\,M_\odot$ and the second broader one at $\sim 1.1\,M_\odot$. The 
average mass of these magnetic white dwarfs is $0.87\,M_\odot$, which is 
higher than the average mass of $0.6\,M_\odot$ for non-magnetic white dwarfs 
(\cite[Tremblay et al. 2019]{tre2019}). A higher incidence
of magnetism in hot massive white dwarfs discovered by the Extreme Ultraviolet 
Explorer (EUVE) had already been reported by \cite[Vennes (1999)]{ven1999}.
There appears to be a lack of cool magnetic white dwarfs with a high mass, 
which is most likely due to their intrinsic faintness
and a shortage of cool white dwarfs in past SDSS colour selections. 

A relationship appears to exist between the mass and magnetic field strength, 
with more massive white dwarfs having stronger magnetic fields, however the 
correlation coefficient is only 0.43 and therefore should be viewed with 
caution. Using population synthesis calculations under the merger scenario for
the generation of fields (see Section 4), \cite[Briggs et al. (2018)]{bri2018} 
predicted that more massive magnetic white dwarfs should have weaker fields 
than less massive magnetic white dwarfs, which contradicts the weak correlation 
observed in Fig.~\ref{fig_mass_teff_dist}. In their simulations they show that
when an asymptotic giant branch (AGB) star merges with a late stage star, it 
produces stronger fields than when an AGB star merges with an earlier main-sequence 
star. However, they also note that double degenerate mergers would not follow 
this trend and instead would result in more massive white dwarfs with stronger 
field strengths and faster rotation than magnetic white dwarfs that
merged in a common envelope phase.

\begin{figure}
    \centering
    \includegraphics[viewport=1 25 565 567,clip, width=0.5\textwidth]{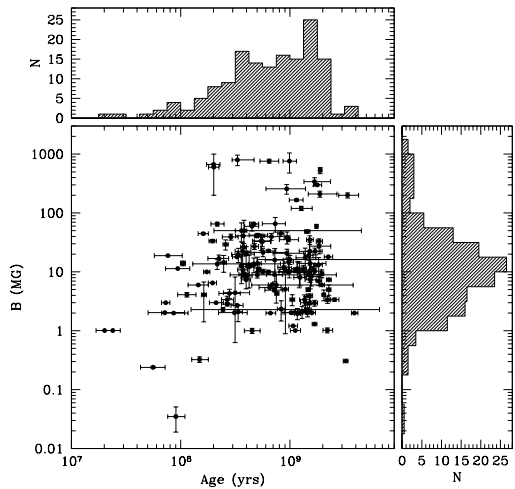}
    \caption{Cooling age versus magnetic field strength distribution of the magnetic DA white dwarfs from SDSS DR12.}
    \label{fig_age}
\end{figure}

Using the SDSS DR2/Gaia sample of magnetic white dwarfs, we can investigate potential correlations between the cooling age of the white 
dwarf and the magnetic field strength and find out if there is any evidence for magnetic field decay or evolution. Fig.~\ref{fig_age} 
shows no obvious correlation between the cooling age versus the magnetic field strength. The cross-correlation
coefficient is only 0.14 indicating that the two properties are not correlated. \cite[Ferrario et al. (2015)]{fer2015} reached the same
conclusions when they compared the magnetic field strength with the effective temperature.

\begin{figure}
    \centering
    \includegraphics[viewport=1 15 565 567,clip, width=0.5\textwidth]{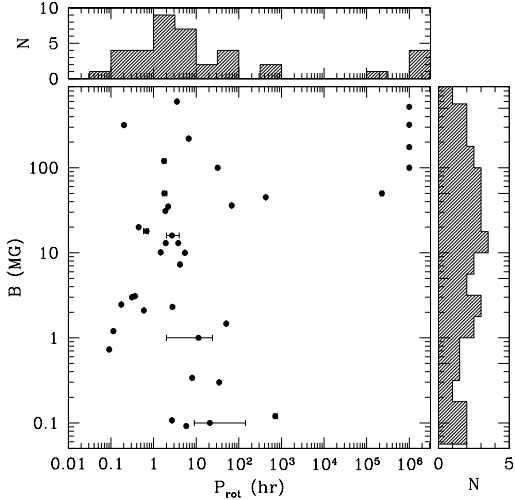}
    \caption{Rotational velocity versus magnetic field strength distribution for known magnetic white dwarfs for which a rotational period
    was measured.}
    \label{fig_rot}
\end{figure}

Many magnetic white dwarfs are faster rotators than their non-magnetic 
counterparts. Most rotational periods in magnetic white dwarfs were 
determined from photometric time series (e.g., \cite[Brinkworth et al. 2013]{bri2013}), although some were discovered using 
polarimetry or spectropolarimetry. Photometric and spectropolarimetric measurements revealed that the massive, high field 
white dwarf RE~J0317-853 (EUVE~J0317-85.5) to be a very fast rotator with a period of 725 seconds
(\cite[Barstow et al. 1995, Ferrario et al. 1997, Vennes et al. 2003]{bar1995,fer1997,ven2003}). It remained the fastest rotating white
dwarf until photometric timeseries of hot DQs revealed them to have rotational periods as short as 5 minutes 
(\cite[e.g., Dunlap et al. 2010, Dufour et al. 2011]{dun2010,duf2011}). Fig.~\ref{fig_rot} shows that most of the magnetic white dwarfs 
have rotation periods shorter than 10 hours with a distribution peaking at 2 - 3 hours. This can be compared to the rotational 
period distribution for non-magnetic white dwarfs determined from pulsation studies showing a mean rotation period of 
35 hours (\cite[Hermes et al. 2017]{her2017}).

Availability of photometric and astrometric data from large scale surveys has 
finally allowed us to better understand the properties of the magnetic white
dwarf population. Detailed, case by case, spectropolarimetric investigations 
remain essential to understand the diverse properties of individual magnetic white dwarfs.

\section{Origin of magnetic fields}

The fossil field origin for the existence of magnetic white dwarfs remains a 
viable model because the magnetic flux of Ap/Bp stars is similar to that of 
magnetic white dwarfs. Assuming flux conservation ($B_{\rm WD}/B_{\rm MS} = (R_{\rm MS}/R_{\rm WD})^2$) 
the magnetic fields of Ap/Bp star can evolve into the strongest fields observed 
in white dwarfs (\cite[Tout et al. 2004]{tou2004}). 
\cite[Angel et al. (1981)]{ang1981} compared the space density of magnetic 
white dwarfs to the space density of magnetic Ap/Bp stars and proposed that 
there is a sufficient number of Ap/Bp stars to generate all known magnetic 
white dwarfs. \cite[Kawka \& Vennes (2004)]{kaw2004} revisited this comparison
and showed that if Ap/Bp stars evolve to high field magnetic white dwarfs 
($B \gtrsim 10$~MG), additional progenitors are required for white dwarfs with 
lower field strengths. \cite[Wickramasinghe \& Ferrario (2005)]{wic2005} 
proposed that white dwarfs with $B \lesssim$~a few ten MG could form from main-sequence stars 
with $M \gtrsim 4.5\,M_\odot$ and low fields of $\approx 10 - 100$~G. 
However, \cite[Auri{\`e}re et al. (2007)]{aur2007} showed that a magnetic 
desert exists below 300 G for A and B type stars, which eliminates
that possibility.

One of the main difficulties with the fossil field origin is that there are no known close, detached magnetic white dwarfs paired 
with a main-sequence star (\cite[Liebert et al. 2015]{lie2015}), although thousands of close, detached but non-magnetic white dwarf 
plus M dwarf spectroscopic binaries have been identified in SDSS \cite[(Rebassa-Mansergas et al. 2016)]{reb2016}.  

To address this conundrum, \cite[Tout et al. (2008)]{tou2008} and 
\cite[Wickramasinghe et al. (2014)]{wic2014} proposed that magnetic fields in 
isolated white dwarfs are created during a common envelope phase experienced 
by the progenitor close binary system.
There are likely variations on the magnetic field generation model proposed by 
\cite[Tout et al. (2008)]{tou2008}. \cite[Potter \& Tout (2010)]{pot2010} and 
\cite[Wickramasinghe et al. (2014)]{wic2014} showed that a dynamo mechanism 
within the common envelope can generate magnetic fields with the strongest 
fields arising in mergers that are differentially rotating near 
break-up.

\cite[Nordhaus et al. (2011)]{nor2011} also proposed a variation on the model,
where the low mass star is tidally disrupted by the pre-white dwarf during the common envelope phase, resulting in the formation of
an accretion 
disc. A dynamo is then generated in the disc creating the magnetic field which is then transferred to the 
degenerate core via accretion.

\cite[Garc{\'\i}a-Berro et al. (2012)]{gar2012} showed that magnetic fields can also be created during the merger
of two white dwarfs. When the white dwarfs merge a hot, convective and differentially rotating corona is created which
produces a dynamo and, thus, the magnetic field. However,
using a population synthesis of binary stars, \cite[Briggs et al. (2015)]{bri2015} showed that the majority of magnetic fields
in white dwarfs would have formed within a common envelope with less than 1\% being formed during the merger of two white dwarfs. 

\cite[Isern et al. (2017)]{ise2017} proposed a model where low magnetic fields ($B \lesssim 0.1$~MG) are generated by phase separation 
when the white dwarf begins to crystallize. They showed that when white dwarfs reach sufficiently low temperatures that the core begins 
crystallizing, phase separation of the main elements occurs. This leads to an unstable, convective liquid mantle on top of a solid core
producing a dynamo which creates a magnetic field.

It is possible that the range of observed magnetic fields stem from several distinct processes. By studying the various 
classes of magnetic white dwarfs,
we may be able to place constraints on the specific processes that could lead to a particular class of magnetic white dwarfs.

\subsection{Double degenerate mergers}

Evidence of magnetic fields being formed in double degenerate mergers has been 
found in common proper motion binaries. EUVE~J0317-85.5 is a hot 
($T_{\rm eff} = 33\,000$~K), strongly magnetic ($B = 450$~MG) and massive 
white dwarf ($M = 1.3\,M_\odot$) in common proper motion with the non-magnetic 
white dwarf LB~9802 (\cite[Barstow et al. 1995, Ferrario et al. 1997, Vennes et al. 2003]{bar1995,fer1997,ven2003}). 
EUVE~J0317-85.5 being significantly more massive than LB~9802 would have had a 
more massive main-sequence progenitor and therefore it should have a much 
longer cooling age compared to LB~9802. However, EUVE~J0317-85.5 is much 
hotter than LB~9802, and therefore its cooling age is also shorter. This 
strongly suggests that EUVE~J0317-85.5 is the result of a merger (\cite[Ferrario et al. 1997]{fer1997}).
Another possible example is the common proper motion binary SDSS~J150746.49+521002.1 plus SDSS~J150746.80+520958.0. SDSS~J1507+5209 is 
magnetic and more massive than SDSS~J1507+5210, however their effective 
temperatures are similar (\cite[Dobbie et al. 2012]{dob2012}). Since they are 
in a common proper motion binary, their total ages should be the same, however 
the total age of the magnetic white dwarf is much shorter than the total age of 
the non-magnetic companion. Therefore SDSS~J1507+5209 is most likely the 
product of a merger and thus the original system was a triple system. 

A small number of common proper motion binaries provide evidence to support 
that some magnetic fields in white dwarfs are formed through mergers. The key 
property is the discrepancy in the total ages of the magnetic white dwarf and 
non-magnetic white dwarf if single star evolution is assumed for both stars.

Spectroscopic follow-up of newly identified common proper motion white dwarf 
pairs may add additional cases and shed more light on the origin of magnetic
fields.

\subsection{Incidence of magnetism}

\begin{table}
  \begin{center}
  \caption{Incidence of magnetism among different classes of white dwarfs.}
  \label{tbl_prototype}
 {\scriptsize
  \begin{tabular}{lcccc}\hline 
{\bf Spectral Type} & {\bf Prototype} & {\bf Reference} & {\bf Fraction (\%)} & {\bf Reference} \\ 
\hline
DAH        & Grw+70$^\circ$8247 & \cite[Kemp et al. (1970)]{kem1970}         & $4\pm1.4$    & \cite[Schmidt \& Smith (1995)]{sch1995} \\ 
DAZH (cool)& G77$-$50           & \cite[Farihi et al. (2011)]{far2011}       & $\sim 50$    & \cite[Kawka et al. (2019)]{kaw2019} \\
DBH        & GD229              & \cite[Swedlund et al. (1974)]{swe1974}     & $\sim 1.5$   & \cite[Kawka (2018)]{kaw2018} \\
DCP        & G195$-$19          & \cite[Angel \& Landstreet (1971)]{ang1971} & $\sim 5$     & \cite[Putney (1997)]{put1997} \\
DZH (cool) & LHS2534            & \cite[Reid et al. (2001)]{rei2001}         & $21.6\pm3.3$ & \cite[Hollands (2017)]{hol2017} \\
DQH (hot)  & SDSS~J1337$+$0026  & \cite[Dufour et al. (2007)]{duf2007}       & $\sim 70$    & \cite[Dufour et al. (2013)]{duf2013} \\
DQH (cool) & G99$-$37           & \cite[Landstreet \& Angel (1971)]{lan1971} & $\sim 4$     & \cite[Vornanen et al. (2013)]{vor2013}     \\
\hline 
  \end{tabular}
  }
 \end{center}
\end{table}

The incidence of magnetism varies between different spectral classes, but it is also affected by the population sampling
criteria. Taking the
population of magnetic white dwarfs as a whole, then in magnitude
limited surveys such as those of Palomar-Green or SDSS 
(\cite[Schmidt \& Smith 1995, Kepler et al. 2013]{sch1995,kep2013}) the incidence is as low as $\sim 5$\%, while 
in volume limited surveys, the incidence is much higher at $13 - 20$\% (\cite[Kawka et al. 2007]{kaw2007}).
Table~\ref{tbl_prototype} lists the many classes of magnetic white dwarfs including the individual prototypes and the
incidence within specific spectral classes.

\subsubsection{Common proper motion binaries}

It was already mentioned that there are no close, detached magnetic white 
dwarfs paired with a main sequence star, however there are a few magnetic 
white dwarfs in common proper motion with a main sequence star. In these 
systems the stars are assumed to have been born at the same time and remain 
distant enough from each other to avoid interaction and thereby follow 
single star evolution. An examination of the incidence of magnetic white dwarfs 
in such systems within the Solar neighbourhood ($d < 20$~pc) reveals that 
magnetic white dwarfs in wide binaries are not that rare. 
\cite[Holberg et al. (2013)]{hol2013} identified 11 so-called Sirius-like 
systems within 20~pc of the Sun. These are systems where the white dwarf is 
paired with a main sequence star with a spectral type of K or earlier. Of 
these 11 binary systems, one has a magnetic white dwarf: WD~1009$-$184 is a 
cool, polluted helium rich white dwarf with a weak surface averaged field of 
$\approx 300$~kG (\cite[Bagnulo \& Landstreet 2019]{bag2019}). Therefore,
this sets an incidence of $\sim 9$\%, comparable to the incidence of single
magnetic white dwarfs. More recently, \cite[Hollands et al. (2018)]{hol2018} 
revisited the 20~pc sample with Gaia and identified a total of 30 white dwarfs 
that are in a wide binary system with a main sequence companion (including M 
dwarfs) and 3 of these are magnetic. These include WD~1009$-$184, along with
the cool DA WD2150$+$591 (\cite[Landstreet \& Bagnulo 2019]{lan2019a}) and the 
cool DQ WD2153$-$512 (\cite[Vornanen et al. 2010]{vor2010}). This corresponds 
to an incidence of $\sim 10$\%. If we exclude systems that are likely to 
interact or have already interacted, then this incidence would be higher. The 
formation of magnetic fields in white dwarfs that are in wide binaries
can be assumed to follow the same paths as those followed by isolated magnetic 
white dwarfs.

\subsubsection{Polluted white dwarfs}

Magnetic fields have been detected in all spectral classes of white dwarfs. 
Most of the known magnetic white dwarfs have hydrogen-rich atmospheres, similar 
to non-magnetic white dwarfs. Some spectral classes of white dwarfs display
significantly higher incidence of magnetism than the general white dwarf 
population. Cool, polluted white dwarfs show an enhanced incidence of magnetism.
\cite[Kawka \& Vennes (2014)]{kaw2014} and \cite[Kawka et al. (2019)]{kaw2019} 
showed that $\approx 50\%$ of cool ($T_{\rm eff} < 6000$~K), polluted 
hydrogen-rich DAZ white dwarfs are magnetic with field strengths below 1~MG. 
Similarly, \cite[Hollands et al. (2015)]{hol2015} and 
\cite[Hollands (2017)]{hol2017} showed that $21.6\pm3.3$\% of cool 
($T_{\rm eff} < 8000$~K) polluted helium-rich DZ white dwarfs are magnetic 
with field strengths ranging from $\approx 0.3$~MG up to 
$\approx 30$~MG. Only two DAZ white dwarfs with an effective temperature above 
6000~K have magnetic fields, NLTT~53908 which is only slightly warmer at 6250~K 
(\cite[Kawka \& Vennes 2014]{kaw2014}) and WD~2105-820 which is much hotter at
$T_{\rm eff} = 10800$~K (\cite[Landstreet et al. 2012]{lan2012}). In the case 
of DZ white dwarfs, \cite[Hollands (2017)]{hol2017} found that there are no 
known magnetic DZ white dwarfs warmer than 8000~K. The lack of warmer 
counterparts to the magnetic, polluted white dwarfs provides constraints on 
the origin of the magnetic field in these stars. Some of these possibilities 
are discussed below.

One possible scenario which may explain the presence of magnetic fields in this particular class involves 
a gaseous planet colliding with an older white dwarf creating differential rotation which would generate a relatively
weak magnetic field (\cite[Farihi et al. 2011, Kawka et al. 2019]{far2011,kaw2019}). The time-scale for a post main sequence 
planetary system to dynamically stabilize is relatively short at about 100 Myr (\cite[Debes \& Sigurdsson 2002]{deb2002}).
Therefore, at the old ages encountered in cool DAZs and DZs ($> 1$~Gyr) a destabilizing agent is required to create a 
planetary or asteroidal debris disc causing atmospheric pollution and contributing to the creation of a magnetic field.
\cite[Farihi et al. (2011)]{far2011} proposed
that planets and asteroids could be destabilized by a close encounter with another star. These cool white dwarfs
have been orbiting the Galaxy for billions of years and \cite[Farihi et al. (2011)]{far2011} estimated that a stellar 
encounter has a 50\% probability of occurring every 0.5 Gyr. Therefore, the older the white dwarf, the more likely it 
will encounter a star that is close enough to perturb the orbit of the outer planets. If one of these planets is a gaseous
type then this would create the magnetic field while the perturbed rocky planets and asteroids would pollute
the white dwarf. The pairing of these events would explain the dual nature of polluted, magnetic white dwarfs.

\cite[Hollands (2017)]{hol2017} suggested that cool, polluted and magnetic 
white dwarfs could have been generated in the cores of giant stars following 
\cite[Kissin \& Thomson (2015)]{kis2015}. In this scenario, a dynamo is 
generated within the giant core during the shell burning phases. They showed 
that angular momentum pumped inward in combination with the absorption of a 
planet or a binary companion could generate a dynamo at the boundary between 
the radiative and convective envelopes. This could create magnetic fields of 
up to about $10^7$ G. However this magnetic field could remain buried below a 
non-magnetized envelope for up to about 1 - 2 Gyr. Since the creation of the 
magnetic field is aided by the engulfment of a planet, it is likely that other 
planets or asteroids could survive and with time migrate closer toward the
white dwarf and be accreted polluting the white dwarf atmosphere.

\subsubsection{Hot DQ white dwarfs}

The class of hot DQs shows the highest incidence of magnetism among the various 
classes of white dwarfs. Approximately 70\% of hot DQs are magnetic 
(\cite[Dufour et al. 2013]{duf2013}). Hot DQs have temperatures ranging from 
18\,000~K up to 24\,000~K and have an atmosphere dominated by carbon 
(\cite[Dufour et al. 2008]{duf2008}). \cite[Coutu et al (2019)]{cou2019} showed 
that they are all massive ($0.8 - 1.2$ M$_\odot$) and about half of these stars have been found to be 
photometrically variable with periods ranging from about 5 minutes up to 
2.1 days. These variations have been attributed to the rotation of the white 
dwarfs (\cite[Dunlap \& Clemens 2015, Williams et al. 2016]{dun2015,wil2016}).
These properties point toward hot DQ white dwarfs being the outcome of mergers 
of white dwarfs. \cite[Coutu et al. (2019)]{cou2019} showed that the hot DQ
sequence extends toward lower temperatures. These cooler objects are also 
massive and they can be distinguished from normal mass ($\sim 0.6$~M$_\odot$) 
DQ white dwarfs by having a higher carbon abundance. 
\cite[Coutu et al. (2019)]{cou2019} also showed that the distribution of transverse 
velocities of these objects is larger than for other classes of objects with 
similar temperatures, suggesting that they must come from an older population. 
If these massive DQs formed in a double degenerate merger, this would provide 
the necessary delay to make these stars appear younger than they really are. 
One star where we clearly see a discrepancy between the total age assuming 
single star evolution and the age based on kinematics is LP~93-21, which is a 
massive DQ with halo like orbit (\cite[Kawka et al. 2020]{kaw2020}) and is
thus highly unlikely to be the unburnt remnant of a subluminous Type Ia 
supernova explosion (\cite[e.g., Vennes et al. 2017]{ven2017}) as proposed
by \cite[Ruffini \& Casey (2019)]{ruf2019}. It is likely that the 
majority of these 
cooler massive DQs will also harbour a magnetic field and be fast rotators. The number of these massive DQs is small compared to 
other classes of white dwarfs, however it is in agreement with the conclusion of \cite[Briggs et al. (2015)]{bri2015} who predicted that
less than about one in four hundred mergers involve double degenerate mergers.

The different properties of the various classes of magnetic white dwarfs, 
along with their respective incidences of magnetism allows links
to be established with potential formation scenarios.

\section{Structure of magnetic atmospheres}

Magnetic fields have the potential to affect the transport of energy within a stellar atmosphere, and therefore the atmospheric structure
as well. Field variations across the stellar surface add more complexities to this problem.

\subsection{Convective energy transport}

As white dwarfs cool, they develop convection zones with a depth increasing 
with age (\cite[Fontaine et al. 2013]{fon2013}). White dwarfs with 
hydrogen-rich atmospheres begin to develop a convection zone 
at about 14\,000~K, whereas white dwarfs with helium-rich atmospheres develop 
convective zones much earlier at about 30\,000~K. As white dwarfs cool,
these convection zones become the main energy transport mechanism, however it is possible that if a magnetic field is present, this
transport of energy can be affected. Suppression of convection by magnetic fields was first investigated by 
\cite[D'Antona \& Mazzitelli (1975)]{dan1975}. Furthermore, \cite[Wickramasinghe \& Martin (1986)]{wic1986} assumed that magnetic
fields can provide resistance to large scale motions and therefore convective energy transport was suppressed in their atmospheric models. 

Recently, \cite[Valyavin et al. (2014)]{val2014} revisited the idea that magnetic fields can suppress convection to explain the higher
incidence of magnetism among cool white dwarfs (\cite[Liebert \& Sion 1979, Valyavin \& Fabrika 1999]{lie1979,val1999}). They argued
that suppression of convection causes a slow down in white dwarf cooling for white dwarfs below 12\,000 to 14\,000~K.
\cite[Tremblay et al. (2015)]{tre2015} conducted radiation magnetohydrodynamic simulations and showed that convection can be 
inhibited in a white dwarf atmosphere by magnetic fields as low as 50~kG, but that the cooling age is not affected until the convective
zone couples with the degenerate core. This occurs at about 6000~K for higher mass white dwarfs ($M\sim1\,M_\odot$) and at lower 
temperatures for lower mass white dwarfs.

Empirical support for suppression of convection was provided by 
\cite[B{\'e}dard, Bergeron \& Fontaine (2017)]{bed2017} and 
\cite[Gentile Fusillo et al. (2018)]{gen2018} using the weakly magnetic white 
dwarf WD~2105$-$820 (\cite[Landstreet et al. 2012]{lan2012}). 
The effective temperature of WD~2105$-$820 is about 10\,000~K and, therefore, 
it is cool enough to develop a shallow convective zone.
\cite[B{\'e}dard, et al. (2017)]{bed2017} showed that using radiative models they were able to match the photometrically and spectroscopically
derived effective temperatures, as well as match the spectroscopically determined distance with the measured parallax. However, convective models
showed discrepancies between the spectroscopically and photometrically determined temperatures, as well as the distance obtained from the
spectroscopic solutions and the measured parallax. Non-magnetic white dwarfs with similar temperature to WD~2105$-$820 show good agreement
between spectroscopically and photometrically determined temperatures when using convective models. Additional support for the 
inhibition of convection was presented by \cite[Gentile Fusillo et al. (2018)]{gen2018}, who showed that consistent results were obtained from fitting ultraviolet and optical spectra when using purely radiative models, whilst fitting with convective models resulted in inconsistent
results. Again, non-magnetic white dwarfs with similar temperatures show consistent results when convective models are used.

Cooler magnetic white dwarfs have deeper and thicker convection zones. \cite[Kawka et al. (2019)]{kaw2019} analyzed the cool 
($\approx 5200$~K), magnetic white dwarf NLTT~7547 using convective and radiative models. Their analysis showed that 
the convective models provide a better fit to the Balmer lines compared to radiative models. They also showed that by suppressing
convection in the model atmospheres, the temperature gradient steepens sharply at larger optical depths and higher pressure. The model
atmospheres also show a downturn in the density at higher pressure which would give rise to a Rayleigh-Taylor instability. They concluded
that convective models are preferred in cool and magnetic white dwarfs. \cite[B{\'e}dard et al. (2017)]{bed2017} also analyzed a sample
of cool (6600 - 8510~K) and magnetic white dwarfs and showed that in these white dwarfs, consistent results between spectroscopic and photometric
solutions, as well as distances inferred from spectroscopy and parallax were 
achieved with convective models. As a result, they concluded
that it may become more difficult to suppress convection in cooler white dwarfs where convective energy transport becomes more important. 

A small number of cool, average mass and magnetic DQs are known 
(\cite[e.g., Schmidt et al. 1999, Vornanen et al. 2010]{sch1999,vor2010}). 
The carbon in the atmospheres of these DQs is dredged up from the core by the 
deep convection zone (\cite[Pelletier et al. 1986]{pel1986}). If convection is 
suppressed by the magnetic field, then we should not find any magnetic DQ white 
dwarfs since there is no convective motion to bring up the carbon to the surface. 
There are two possibilities to explain the existence of these cool and magnetic 
DQs. The first is that the carbon is not dredged up but it has been present in 
the atmosphere from the outset and as such they could be the descendants of 
hot DQs. In this case, they should also be massive. The two prototypical 
magnetic DQs, LHS2229 and LP~790-29, have $T_{\rm eff} < 5000$~K 
and $M \lesssim 0.7\,M_\odot$ (\cite[Blouin et al. 2019]{blo2019}). They are also not known to vary rapidly. The second explanation is the 
same as for the cool magnetic DA white dwarfs, that is the convection zone is too deep and therefore the magnetic field cannot suppress it.

\subsection{Magnetic field structure}

The magnetic field structure in a white dwarf can provide clues to how the 
magnetic field was created. Complex magnetic fields are predicted for white 
dwarfs that formed in a merger (\cite[Garc{\'\i}a-Berro, et al. 2012]{gar2012}).
Simpler fields that are approximately dipolar maybe an attribute of white 
dwarfs that are descendants of Ap or Bp stars 
(\cite[Braithwaite \& Spruit 2004]{bra2004}). When modelling the magnetic field 
geometry in white dwarfs, a centred or offset dipole is often assumed. However, 
rotating white dwarfs
provide the means to study the field structure and reveal a diversity in the field geometry. Depending on the magnetic field structure
the Zeeman splitted lines may vary as a function of the rotational phase and reveal how the magnetic field strength varies on the white dwarf
surface. In addition to measuring the strength of the magnetic field, circular and linear polarization can provide the direction of the magnetic
field. With a series of circular polarization spectra covering the rotation period of 0.243 days, \cite[Landstreet et al. (2017)]{lan2017} 
showed that WD~2047$+$372 can be modelled by a simple dipole with a weak field of $91.8\pm0.8$~kG. However, their analysis of WD~2359$-$434 
revealed a more complex field structure, which required a combination of a 
dipolar field and a non-aligned quadrupolar field.

Detailed modelling of rotating white dwarfs with stronger fields has resulted in even more complex structures. The high field, massive
and hot white dwarf EUVE~J0317$-$855 has a field of 185~MG with a magnetic spot of 425~MG 
(\cite[Burleigh et al. 1999, Vennes et al. 2003]{bur1999,ven2003}). Another white dwarf with a magnetic spot is WD~1953-011 
(\cite[Maxted et al. 2000, Valyavin et al. 2008]{max2000,val2008}) which was revealed over its rotational period of 1.4418 days 
(\cite[Brinkworth et al. 2005]{bri2005}). \cite[Euchner et al. (2006)]{euc2006} used Zeeman tomography to show that the magnetic field of
PG~1015$+$014 is a combination of three individually offset non-aligned dipoles.

Our study of the structure of magnetic white dwarf atmospheres offers 
computational challenges. Complete suppression of the convection zone appears 
unlikely at lower temperatures. Complex surface field distributions are 
observed in several white dwarfs but such observations are challenging and 
impractical in most cases.

\section{Summary}

With the growth of large scale photometric, spectroscopic and astrometric surveys we have been able to study the magnetic white dwarf population 
as a whole in greater detail. The growing sample of white dwarfs has revealed that the incidence of magnetism varies 
across the spectral classes, with the class of hot DQs showing
the highest incidence at about 70\%. There is evidence that these objects are a result of double degenerate mergers, which is
one of the proposed scenarios for the creation of magnetic fields.

There are several proposed scenarios for the creation of magnetic fields in white dwarfs, with the merging scenario being
favoured, although many challenges remain. It is also likely that magnetic fields in white dwarfs may be created by several distinct processes.

The magnetic field structure across the surface can be complex. We have some observational evidence that shows that the white dwarf
atmospheric structure is affected by the presence of a magnetic field, such as the suppression of convection. This provides
computational challenges in modelling white dwarf atmospheres in the presence of complex magnetic fields.

\subsubsection*{Acknowledgements} AK thanks S. Vennes, L. Ferrario and D.T. Wickramasinghe for stimulating discussions and the Science
Organizing Committee for the kind invitation to present this review. The International Centre for Radio Astronomy Research is a joint venture between Curtin University and the University of Western Australia, funded by the state
government of Western Australia and the joint venture partners.


\begin{thebibliography}{}

\bibitem[Alam, et al. (2015)]{ala2015} {Alam, S., et al.} 2015, \textit{ApJS}, 219, 12
\bibitem[Al-Hujaj \& Schmelcher (2003)]{alh2003} {Al-Hujaj, O.-A., Schmelcher, P.} 2003, \textit{PhRvA}, 68, 053403
\bibitem[Angel, Borra \& Landstreet (1981)]{ang1981} {Angel, J.~R.~P., Borra, E.~F., Landstreet, J.~D.} 1981, \textit{ApJS}, 45, 457
\bibitem[Angel \& Landstreet (1971)]{ang1971} {Angel, J.~R.~P., Landstreet, J.~D.} 1971, \textit{ApJ} Letters, 164, L15
\bibitem[Angel \& Landstreet (1974)]{ang1974} {Angel, J.~R.~P., Landstreet, J.~D.} 1974, \textit{ApJ}, 191, 457
\bibitem[Appenzeller, et al. (1998)]{app1998} {Appenzeller, I., et al.} 1998, \textit{Msngr}, 94, 1
\bibitem[Auri{\`e}re, et al. (2007)]{aur2007} {Auri{\`e}re M., et al.} 2007, \textit{A\&A}, 475, 1053
\bibitem[Babcock (1947)]{bab1947} {Babcock, H.~W.} 1947, \textit{ApJ}, 105 105
\bibitem[Bagnulo \& Landstreet (2019)]{bag2019} {Bagnulo, S., Landstreet, J.~D.} 2019, \textit{A\&A}, 630, A65
\bibitem[Barstow, et al. (1995)]{bar1995} {Barstow, M.~A., Jordan, S., O'Donoghue, D., Burleigh, M.~R., Napiwotzki, R., Harrop-Allin, M.~K.} 1995, \textit{MNRAS}, 277, 971
\bibitem[Becken \& Schmelcher (2001)]{bec2001} {Becken, W., Schmelcher, P.} 2001, \textit{PhRvA}, 63, 053412
\bibitem[B{\'e}dard, Bergeron \& Fontaine (2017)]{bed2017} {B{\'e}dard, A., Bergeron, P., Fontaine, G.} 2017, \textit{ApJ}, 848, 11
\bibitem[Benvenuto \& Althaus (1999)]{ben1999} {Benvenuto, O.~G., Althaus, L.~G.} 1999, \textit{MNRAS}, 303, 30
\bibitem[Berdyugina, Berdyugin \& Piirola (2007)]{ber2007} {Berdyugina, S.~V., Berdyugin, A.~V., Piirola, V.} 2007, \textit{PhRvL}, 99, 091101
\bibitem[Blackett (1947)]{bla1947} {Blackett, P.~M.~S.} 1947, \textit{Nature}, 159, 658
\bibitem[Blouin, et al. (2019)]{blo2019} {Blouin, S., Dufour, P., Thibeault, C., Allard, N.~F.} 2019, \textit{ApJ}, 878, 63
\bibitem[Braithwaite \& Spruit (2004)]{bra2004} {Braithwaite, J., Spruit, H.~C.} 2004, \textit{Nature}, 431, 819
\bibitem[Briggs, et al.(2015)]{bri2015} {Briggs, G.~P., Ferrario, L., Tout, C.~A., Wickramasinghe, D.~T., Hurley, J.~R.} 2015, \textit{MNRAS}, 447, 1713
\bibitem[Briggs, et al. (2015)]{bri2018} {Briggs, G.~P., Ferrario, L., Tout, C.~A., Wickramasinghe, D.~T.} 2018, \textit{MNRAS}, 478, 899
\bibitem[Brinkworth, et al. (2013)]{bri2013} {Brinkworth, C.~S., Burleigh, M.~R., Lawrie, K., Marsh, T.~R., Knigge, C.} 2013, \textit{ApJ}, 773, 47
\bibitem[Brinkworth, et al. (2005)]{bri2005} {Brinkworth, C.~S., Marsh, T.~R., Morales-Rueda, L., Maxted, P.~F.~L., Burleigh, M.~R., Good, S.~A.} 2005, \textit{MNRAS}, 357, 333
\bibitem[Burleigh, Jordan \& Schweizer (1999)]{bur1999} {Burleigh, M.~R., Jordan, S., Schweizer, W.} 1999, \textit{ApJ} Letters, 510, L37
\bibitem[Coutu, et al. (2019)]{cou2019} {Coutu, S., Dufour, P., Bergeron, P., Blouin, S., Loranger, E., Allard, N.~F., Dunlap, B.~H.} 2019, \textit{ApJ}, 885, 74
\bibitem[D'Antona \& Mazzitelli (1975)]{dan1975} {D'Antona, F., Mazzitelli, I.} 1975, \textit{A\&A}, 42, 127
\bibitem[Dobbie, et al. (2012)]{dob2012} {Dobbie, P.~D., et al.} 2012, \textit{MNRAS}, 421, 202
\bibitem[Donati, et al. (2006)]{don2006} {Donati, J.-F., Catala, C., Landstreet, J.~D., Petit, P.} 2006, ASP Conf. Ser. Vol. 358, 362 \bibitem[Dufour, et al. (2007)]{duf2007} {Dufour, P., Liebert, J., Fontaine, G., Behara, N.} 2007, \textit{Nature}, 450, 522
\bibitem[Dufour, et al. (2011)]{duf2011} {Dufour, P., B\'eland, S., Fontaine, G., Chayer, P., Bergeron, P.} 2011, \textit{ApJ} Letters, 733, L19
\bibitem[Dufour, et al. (2008)]{duf2008} {Dufour, P., Fontaine, G., Liebert, J., Schmidt, G.~D., Behara, N.} 2008, \textit{ApJ}, 683, 978
\bibitem[Dufour, et al. (2013)]{duf2013} {Dufour, P., Vornanen, T., Bergeron, P., Fontaine, G., Berdyugin, A.} 2013, ASP Conf. Ser. Vol. 469, 167
\bibitem[Dunlap, Barlow \& Clemens (2010)]{dun2010} {Dunlap, B.~H., Barlow, B.~N., Clemens, J.~C.} 2010, \textit{ApJ} Letters, 720, L159
\bibitem[Dunlap \& Clemens (2015)]{dun2015} {Dunlap, B.~H., Clemens, J.~C.} 2015, ASP Conf. Ser., Vol. 493, 547
\bibitem[Euchner, et al. (2006)]{euc2006} {Euchner, F., Jordan, S., Beuermann, K., Reinsch, K., G{\"a}nsicke, B.~T.} 2006, \textit{A\&A}, 451, 671
\bibitem[Farihi et al. (2011)]{far2011} {Farihi, J., Dufour, P., Napiwotzki, R., Koester, D.} 2011, \textit{MNRAS}, 413, 2559
\bibitem[Ferrario, de Martino \& G\"{a}nsicke (2015)]{fer2015} {Ferrario, L., de Martino, D., G\"{a}nsicke, B.~T.} 2015, \textit{SSRv}, 191, 111
\bibitem[Ferrario, et al. (1997)]{fer1997} {Ferrario, L., Vennes, S., Wickramasinghe, D.~T., Bailey, J.~A., Christian, D.~J.} 1997, \textit{MNRAS}, 292, 205
\bibitem[Ferrario, Wickramasinghe \& Kawka (2019)]{fer2019} {Ferrario, L., Wickramasinghe, D., Kawka, A.} 2019, \textit{AdSpR}, in press 
\bibitem[Fontaine, et al. (2013)]{fon2013} {Fontaine, G., Brassard, P., Charpinet, S., Randall, S.~K., Van Grootel, V.} 2013, ASP Conf. Ser. Vol. 479, 211
\bibitem[Garc{\'\i}a-Berro, et al. (2012)]{gar2012} {Garc{\'\i}a-Berro E., et al.} 2012, \textit{ApJ}, 749, 25
\bibitem[Gentile Fusillo, et al. (2018)]{gen2018} {Gentile Fusillo, N.~P., Tremblay, P.-E., Jordan, S., G{\"a}nsicke, B.~T., Kalirai, J.~S., Cummings, J.} 2018, \textit{MNRAS}, 473, 3693
\bibitem[Gonz{\'a}lez-F{\'e}rez \& Schmelcher (2003)]{gon2003} {Gonz{\'a}lez-F{\'e}rez, R., Schmelcher, P.} 2003, \textit{EPJD}, 23, 189
\bibitem[Greenstein (1969)]{gre1969} {Greenstein, J.~L.} 1969, \textit{ApJ}, 158, 281
\bibitem[Hermes, et al. (2017)]{her2017} {Hermes, J.~J., et al.} 2017, \textit{ApJS}, 232, 23
\bibitem[Herzberg (1945)]{her1945} {Herzberg, G.} 1945, \textit{Atomic Spectra and Atomic Structure}, New York: Dover
\bibitem[Holberg, et al. (2013)]{hol2013} {Holberg, J.~B., Oswalt, T.~D., Sion, E.~M., Barstow, M.~A., Burleigh, M.~R.} 2013, \textit{MNRAS}, 435, 2077
\bibitem[Hollands (2017)]{hol2017} {Hollands, M.~A.} 2017, PhD Thesis, Univ. Warwick
\bibitem[Hollands, et al. (2015)]{hol2015} {Hollands, M.~A., G{\"a}nsicke, B.~T., Koester, D.} 2015, \textit{MNRAS}, 450, 681
\bibitem[Hollands, et al. (2018)]{hol2018} {Hollands, M.~A., Tremblay, P.-E., G{\"a}nsicke, B.~T., Gentile-Fusillo, N.~P., Toonen, S.} 2018, \textit{MNRAS}, 480, 3942
\bibitem[Isern, et al. (2017)]{ise2017} {Isern, J., Garc{\'\i}a-Berro, E., K{\"u}lebi, B., Lor{\'e}n-Aguilar, P.} 2017, \textit{ApJ} Letters, 836, L28
\bibitem[Jordan, et al. (1998)]{jor1998} {Jordan, S., Schmelcher, P., Becken, W., Schweizer, W.} 1998, \textit{A\&A}, 336, L33
\bibitem[Kawka (2018)]{kaw2018} {Kawka, A.} 2018, CoSka, 48, 228
\bibitem[Kawka \& Vennes (2004)]{kaw2004} {Kawka, A., Vennes, S.} 2004, Proc. IAU Symp. 224, 879
\bibitem[Kawka \& Vennes (2012)]{kaw2012} {Kawka, A., Vennes, S.} 2012, \textit{MNRAS}, 425, 1394
\bibitem[Kawka \& Vennes (2014)]{kaw2014} {Kawka, A., Vennes, S.} 2014, \textit{MNRAS}, 439, L90
\bibitem[Kawka, Vennes \& Ferrario (2020)]{kaw2020} {Kawka, A., Vennes, S., Ferrario, L.} 2020, \textit{MNRAS}, 491, L40
\bibitem[Kawka, et al. (2019)]{kaw2019} {Kawka, A., Vennes, S., Ferrario, L., Paunzen, E.} 2019, \textit{MNRAS}, 482, 5101
\bibitem[Kawka, et al. (2007)]{kaw2007} {Kawka, A., Vennes, S., Schmidt, G.~D., Wickramasinghe, D.~T., Koch, R.}
2007, \textit{ApJ}, 654, 499
\bibitem[Kemic (1974)]{kem1974} {Kemic, S.~B.} 1974, \textit{JILA Pub.} 1154
\bibitem[Kemic (1975)]{kem1975} {Kemic, S.~B.} 1975, \textit{Ap\&SS}, 36, 459
\bibitem[Kemp et al. (1970)]{kem1995} {Kemp, J.~C., Swedlund, J.~B., Landstreet, J.~D., Angel, J.~R.~P.} 1970,
\textit{ApJ} (Letters), 161, L77
\bibitem[Kepler, et al.(2013)]{kep2013} {Kepler S.~O., et al.} 2013, \textit{MNRAS}, 429, 2934
\bibitem[Kissin \& Thompson (2015)]{kis2015} Kissin, Y., Thompson, C., 2015, \textit{ApJ}, 809, 108
\bibitem[Landi Degl'Innocenti \& Landolfi (2004)]{lan2004} Landi Degl'Innocenti E., Landolfi M., 2004, Polarization in spectral lines, Vol. 307, om Astrophysics and Space Library, Kluwer Academic Publishers, Dordrecht
\bibitem[Landstreet \& Angel (1971)]{lan1971} {Landstreet, J.~D., Angel, J.~R.~P.} 1971, \textit{ApJ} Letters, 165, L67
\bibitem[Landstreet \& Bagnulo (2019)]{lan2019a} {Landstreet, J.~D., Bagnulo, S.} 2019, \textit{A\&A}, 623, A46
\bibitem[Landstreet \& Bagnulo (2019)]{lan2019b} {Landstreet, J.~D., Bagnulo, S.} 2019, \textit{A\&A}, 628, A1
\bibitem[Landstreet, et al. (2012)]{lan2012} {Landstreet, J.~D., Bagnulo, S., Valyavin, G.~G., Fossati, L., Jordan, S., Monin, D., Wade, G.~A.} 2012, \textit{A\&A}, 545, A30
\bibitem[Landstreet, et al. (2017)]{lan2017} {Landstreet, J.~D., Bagnulo, S., Valyavin, G., Valeev, A.~F.} 2017, \textit{A\&A}, 607, A92
\bibitem[Liebert, et al. (1978)]{lie1978} {Liebert, J., Angel, J.~R.~P., Stockman, H.~S., Beaver, E.~A.} 1978, \textit{ApJ}, 225, 181
\bibitem[Liebert, et al. (2015)]{lie2015} {Liebert, J., Ferrario, L., Wickramasinghe, D.~T., Smith, P.~S.} 2015,
\textit{ApJ}, 804, 93
\bibitem[Liebert \& Sion (1979)]{lie1979} {Liebert, J., Sion, E.~M.} 1979, \textit{ApL}, 20, 53
\bibitem[Lindegren, et al. (2018)]{lin2018} {Lindegren, L., et al.} 2018, \textit{A\&A}, 616, A2
\bibitem[Maxted, et al. (2000)]{max2000} {Maxted, P.~F.~L., Ferrario, L., Marsh, T.~R., Wickramasinghe, D.~T.} 2000, \textit{MNRAS}, 315, L41
\bibitem[Nordhaus, et al. (2011)]{nor2011} {Nordhaus, J., Wellons, S., Spiegel, D.~S., Metzger B.~D., Blackman E.~G.} 2011, \textit{PNAS}, 108, 3135
\bibitem[Pelletier, et al. (1986)]{pel1986} {Pelletier, C., Fontaine, G., Wesemael, F., Michaud, G., Wegner, G.} 1986, \textit{ApJ}, 307, 242
\bibitem[Potter \& Tout (2010)]{pot2010} {Potter, A.~T., Tout, C.~A.} 2010, \textit{MNRAS}, 402, 1072
\bibitem[Putney (1997)]{put1997} {Putney A.} 1997, \textit{ApJS}, 112, 527
\bibitem[Rebassa-Mansergas, et al. (2016)]{reb2016} {Rebassa-Mansergas, A., et al.} 2016, \textit{MNRAS}, 458, 3808
\bibitem[Reid, Liebert \& Schmidt (2001)]{rei2001} {Reid, I.~N., Liebert, J., Schmidt, G.~D.} 2001, \textit{ApJ}Letters, 550, L61
\bibitem[Ruffini \& Casey (2019)]{ruf2019} {Ruffini, N.~J., Casey, A.~R.} 2019, \textit{MNRAS}, 489, 420
\bibitem[Ryabchikova, et al. (2015)]{rya2015} {Ryabchikova, T., Piskunov, N., Kurucz, R.~L., Stempels, H.~C., Heiter, U., Pakhomov, Y., Barklem, P.~S.} 2015, \textit{PhyS}, 90, 054005
\bibitem[Schimeczek \& Wunner (2014)]{sch2014} {Schimeczek, C., Wunner, G.} 2014, \textit{ApJS}, 212, 26
\bibitem[Schmidt \& Smith (1994)]{sch1994} {Schmidt, G.~D., Smith, P.~S.} 1994, 
\textit{ApJ} (Letters), 423, L63
\bibitem[Schmidt \& Smith (1995)]{sch1995} {Schmidt, G.~D., Smith, P.~S.} 1995, \textit{ApJ}, 448, 305
\bibitem[Schmidt, et al. (1999)]{sch1999} {Schmidt, G.~D., Liebert, J., Harris, H.~C., Dahn, C.~C., Leggett, S.~K.} 1999, \textit{ApJ}, 512, 916
\bibitem[Swedlund, et al. (1974)]{swe1974} {Swedlund, J.~B., Wolstencroft, R.~D., Michalsky, J.~J., Kemp, J.~C.} 1974, \textit{ApJ} Letters, 187, L121
\bibitem[Tout, Wickramasinghe \& Ferrario (2004)]{tou2004} {Tout, C.~A., Wickramasinghe, D.~T., Ferrario, L.} 2004, \textit{MNRAS}, 355, L13
\bibitem[Tout, et al. (2008)]{tou2008} {Tout, C.~A., Wickramasinghe, D.~T., Liebert, J., Ferrario, L., Pringle, J.~E.} 2008, \textit{MNRAS}, 387, 897
\bibitem[Tremblay, et al. (2019)]{tre2019} {Tremblay, P.-E., Cukanovaite, E., Gentile Fusillo, N.~P., Cunningham, T., Hollands, M.~A.} 2019, \textit{MNRAS}, 482, 5222
\bibitem[Tremblay, et al. (2015)]{tre2015} {Tremblay, P.-E., et al.} 2015, \textit{ApJ}, 812, 19
\bibitem[Valyavin \& Fabrika (1999)]{val1999} {Valyavin, G., Fabrika, S.} 1999, ASP Conf. Ser. Vol. 169, 206
\bibitem[Valyavin, et al. (2008)]{val2008} {Valyavin, G., Wade, G.~A., Bagnulo, S., Szeifert, T., Landstreet, J.~D., Han, I., Burenkov, A.} 2008, \textit{ApJ}, 683, 466
\bibitem[Valyavin, et al. (2014)]{val2014} {Valyavin, G., et al.} 2014, \textit{Nature}, 515, 88
\bibitem[Vennes (1999)]{ven1999} {Vennes, S.} 1999, \textit{ApJ}, 525, 995
\bibitem[Vennes, et al. (2003)]{ven2003} {Vennes, S., Schmidt, G.~D., Ferrario, L., Christian, D.~J., Wickramasinghe, D.~T., Kawka, A.} 2003, \textit{ApJ}, 593, 1040
\bibitem[Vennes, et al. (2017)]{ven2017} {Vennes, S., Nemeth, P., Kawka, A., Thorstensen, J.~R., Khalack, V., Ferrario, L., Alper, E.~H.} 2017, \textit{Science}, 357, 680
\bibitem[Vornanen, et al. (2010)]{vor2010} {Vornanen, T., Berdyugina, S.~V., Berdyugin, A.~V., Piirola, V.} 2010, \textit{ApJ} Letters, 720, L52
\bibitem[Vornanen, Berdyugina \& Berdyugin (2013)]{vor2013} {Vornanen, T., Berdyugina, S.~V., Berdyugin, A.} 2013, \textit{A\&A}, 557, A38
\bibitem[Wickramasinghe \& Ferrario (2005)]{wic2005} {Wickramasinghe, D.~T., Ferrario, L.} 2005, \textit{MNRAS}, 356, 1576
\bibitem[Wickramasinghe \& Martin (1986)]{wic1986} {Wickramasinghe, D.~T., Martin, B.} 1986, \textit{MNRAS}, 223, 323
\bibitem[Wickramasinghe, et al. (2002)]{wic2002} {Wickramasinghe, D.~T., Schmidt, G., Ferrario, L., Vennes, S.} 2002, \textit{MNRAS}, 332, 29
\bibitem[Wickramasinghe, Tout \& Ferrario (2014)]{wic2014} {Wickramasinghe, D.~T., Tout, C.~A., Ferrario, L.} 2014, \textit{MNRAS}, 437, 675
\bibitem[Williams, et al. (2016)]{wil2016} {Williams, K.~A., Montgomery, M.~H., Winget, D.~E., Falcon, R.~E., Bierwagen, M.} 2016, \textit{ApJ}, 817, 27
\end{thebibliography}
\end{document}